\documentclass[showpacs,showkeys,amsmath,amssymb,preprint,nofootinbib]{revtex4}
\usepackage{graphicx}
\usepackage{epsfig}
\usepackage{bm}
\usepackage{float}
\usepackage{fontenc}
\usepackage{graphicx}
\usepackage{bm}
\usepackage{amsmath}
\def\beq{\begin{eqnarray}}

\def\eeq{\end{eqnarray}}

\begin{document}

\title[DM-DE]{Holographic approach for dark energy - dark matter interaction in curved FLRW spacetime}

\author{Miguel Cruz}
\email{miguelcruz02@uv.mx}
\affiliation{Facultad de F\'\i sica, Universidad Veracruzana 91000, Xalapa, Veracruz, M\'exico}

\author{Samuel Lepe}
\email{samuel.lepe@pucv.cl}
\affiliation{Instituto de F\'\i sica, Pontificia Universidad Cat\'olica de Valpara\'\i so, Casilla 4950, Valpara\'\i so, Chile}

\date{\today}

\begin{abstract}
In this work we explore under a holographic approach the dark energy - dark matter interaction in a non-flat Friedmann-Lemaitre-Robertson-Walker spacetime, with a cut-off for the dark energy component given in terms of the Hubble scale. Based on results coming from the use of observational data, we consider a positive interaction $Q$-term together with a Chevallier-Polarsky-Linder type parametrization for the coincidence parameter, we realize that the model admits a Type III future singularity, in this singular universe we have that for a null value of the curvature parameter its scale factor is a constant value. With the use of some cosmological parameters constrained with observational data, we obtain that the crossing of the phantom divide is possible. Once the singular nature is obtained for the model, its Statefinder diagnosis reveals a behavior that is {\it far} from the $\Lambda$-CDM model. 
\end{abstract}
\keywords{holography, dark energy, singularity}

\pacs{95.36.+x, 95.35.+d, 98.80.-k}

\maketitle

\section{Introduction}
It is a well established fact that the actual stage of the universe presents an accelerated expansion \cite{riess, perlmutter, tegmark, observations, observations1} and it is believed that an {\it exotic} component of the universe named {\it dark energy} is responsible for such expansion. Several models have been proposed to explain this unknown component, however our knowledge is still limited, specifically when some questions about the underlying mechanisms of the past and future evolution of the universe arise. Moreover, until now there is no a conclusive unified scenario for the early and late universe. See for instance the Refs. \cite{tsujikawa, Q1} for a general perspective on the subject. In a complementary manner, in Refs. \cite{ref, ref1, ref2} we can find some interesting proposals to try to understand the current cosmic acceleration, these are based on modifications to the theory of gravity, an alternative description of gravitational theory through a general function of the torsion scalar and quintom cosmology.\\

Despite the existence of distinct promising models for dark energy, we must take into account that their validity is dictated by the observations. It is at this point that the word {\it exotic} becomes meaningful because the prevailing tendency shown by the observational data is that the cosmic component driving the current accelerated expansion has a negative equation of state parameter (or simply $\omega$-parameter) and can reach values ​​less than -1 (for the latest results the Ref. \cite{des} can be seen), that is, the crossing to the so-called phantom zone is allowed. In reference \cite{caldwell} it was shown for the first time that this peculiar characteristic is not ruled out by observational data. Subsequently with the use of supernova observations the crossing was confirmed within the framework of quintom cosmology \cite{quintom}. Besides, in Refs. \cite{quintom2, quintom3} can be found that the crossing was obtained for distincts models of quintom cosmology. It is believed that once the universe enters to the phantom zone, its final state will be given by a singularity where all matter and spacetime itself disintegrates, however, although the crossing to the phantom zone may be possible, was found that within the quintom cosmology scenario the universe could have an oscillating behavior \cite{osci1, osci2}. It is important to point out that there is a classification that shows us different types of future singularities, in this work we could determine that in a holographic scenario the final destiny for a curved universe will be given by a singularity of Type III, we will discuss this in more detail later. The aforementioned stage for the universe has led to the origin of various proposals among which we can highlight the scalar field approach, however by making the crossing to the phantom zone possible, the model is not longer consistent with the dominant energy condition and exhibits certain instabilities \cite{sf}. Recently it was possible to demonstrate that the phantom-quintessence scenario can also take place when dissipative effects are considered in a causal thermodynamics scheme \cite{us, us2, us3}. Additionally, within this scheme it is possible to determine that a scenario with the presence of a future singularity at a finite time can emerge naturally due only to bulk viscosity effects \cite{brevik} and it is known that this scenario represents a realistic alternative to the $\Lambda$-CDM cosmology \cite{frampton}.\\

An auspicious scenario that currently exists to try to explain the nature of dark energy is the one in which its interaction with the dark matter is considered through a $Q$-term. This composition is also referred to as {\it dark cosmological sector} \cite{SdelC}. Within this framework, the presence of these interactions could lead to a more realistic scenario for the universe. It is important to point out that the interacting scheme has been widely studied since can be supported by the observational data \cite{Q1, Qdiv, arevalo} and additionally was found that every future singularity induced in this kind of model can be mapped into a singular behavior of the $Q$-interaction term, this means that the energy flow between the dark energy - dark matter sectors will diverge \cite{Qdiv}. Recent results provided by the EDGES collaboration reveal some discrepancies between the signal observed and the predictions for the 21-cm hyperfine transition spectrum, however, in Refs. \cite{211, 212} was shown that under the scheme of interacting dark energy the aforementioned discrepancies can be explained. In Ref. \cite{aydiner} it was found that the interaction between these components can be modeled by using some kind of non-linear Lotka-Volterra equations adapted for cosmology, obtaining as principal result a chaotic universe. In this form the Big-Bang or oscillating universe ideas can be carried out by this single model, this new perspective of the interacting approach seems to give a few clues that could help to solve some important cosmological problems.\\

In this work we will consider a holographic approach for the dark energy interacting scheme as an alternative to describe the current accelerated expansion of the universe and we will also explore the possibility of a future singularity (phantom scenario) within this framework when the effects of spatial curvature of spacetime are included. As we will see, the curvature parameter among other cosmological parameters of the model appear in the constructed interaction term. A similar approach can be found in Ref. \cite{closed} for a closed universe but with the difference that in this description the form of the interaction term is given a priori.
In Refs. \cite{st1, st2} was shown that a realistic unified scenario for an early-time inflationary and late-time accelerating phantom universe can be obtained by considering a holographic dark energy within some scalar-tensor theories and the $f(R)$ gravity model.\\ 

It is important to mention that late results coming from constraining the Hubble parameter for several models, showed that the $\Lambda$-CDM model does not rule out non-flat models or dynamical dark energy models, but better results are obtained for those models in which both components are allowed \cite{omega1, omega}. In fact, the latest results coming from the Planck collaboration do not discard an universe with non-zero spatial curvature, see Ref. \cite{observations} where can be found the curvature parameter given as $\Omega _{k}\left( 0\right)
= 0.000_{-0.005\left(k = 1\right)}^{+0.005(k = -1)}$.\\

The paper is organized as follows: In Sect. (\ref{sec:IDEDM}) we provide the dynamics of the model in non-flat FLRW spacetime under the interacting approach and we write some quantities of interest at cosmological level such as the coincidence and deceleration parameter. We also consider a specific cut-off for the dark energy component. Using some recent observational data we determine the range of values for each cosmological parameter up to the present time, we focus on the equation of state parameter. In Sect. (\ref{sec:future}) we consider a Chevallier-Polarsky-Linder (CPL)-type parametrization for the coincidence parameter and determine that under this description, the model admits a Type III future singularity and we point out some of its characteristics. Once the singularity is identified, we calculate the form of cosmological parameters of the model with the presence of the singularity and state the general behavior of these cosmological parameters near the singularity, up to the present and for the early universe. In Sect. (\ref{sec:state}) we perform a Statefinder diagnosis in order to characterize this dark energy model, as a result we show that the model is {\it far} from the cosmological standard model. Finally, in Sect. (\ref{sec:final}) we write the conclusions of our work.

\section{Interacting dark matter-dark energy scheme}
\label{sec:IDEDM}
In this section we will describe briefly the dynamics of the interacting scheme for the dark matter - dark energy components. We show that under the election of a cut-off given by the Hubble scale for the dark energy density, $\rho_{DE}$, we can construct a specific interaction term between the aforementioned components, which results to be a function of the parameters of the model and the cosmological redshift. We comment about the positivity (negativity) of the interaction term obtained at the end of the section.\\ 
In the non-flat FLRW spacetime the Friedmann constraint can be written as follows
\begin{equation}
E^{2}(z) = \frac{1}{3H^{2}_{0}}\left(\rho_{DE}(z)+\rho_{DM}(z)\right) + \Omega_{k}(z),
\label{eq:fried} 
\end{equation} 
where $E(z) = H(z)/H_{0}$ is the normalized Hubble parameter, $z$ is the redshift which is defined through the relation $1+z = a_{0}/a$, $a$ is the cosmic scale factor\footnote{With the subscript $0$ we mean that the cosmological parameters are evaluated at present time ($z=0$).}, $\rho_{DE}$ and $\rho_{DM}$ are the energy densities for dark energy and dark matter. Besides, $\Omega_{k}$ is the curvature parameter defined as $\Omega_{k}(z) = \Omega_{k}(0)(1+z)^{2}$ where $\Omega_{k}(0) = - k/a^{2}_{0}H^{2}_{0}$, being $k$ the parameter that characterizes the topology of the spacetime, $k = \pm 1, 0$ for a closed, open and flat universe, respectively.\\ 
The continuity equations for the energy densities are given by
\begin{align}
& \rho'_{DE} - 3\left(\frac{1+\omega_{DE}}{1+z} \right)\rho_{DE} = \frac{Q}{H_{0}E(z)(1+z)},\label{eq:cont1}\\
& \rho'_{DM} - \left(\frac{3}{1+z} \right)\rho_{DM} = -\frac{Q}{H_{0}E(z)(1+z)},
\label{eq:cont2}
\end{align}  
where $\omega = p/\rho$ is the equation of state parameter, we have assumed $\omega_{DM} = 0$. The prime denotes derivative with respect to the redshift. On the other hand, the $Q$-terms determine the behavior of the interaction between the dark energy and dark matter. By using the Eqs. (\ref{eq:fried}), (\ref{eq:cont1}) and (\ref{eq:cont2}) one gets
\begin{equation}
1+\frac{\omega_{DE}(z)}{1+r(z)} = \frac{2}{3}\left(\frac{1}{2}(1+z)\frac{d \ln E^{2}(z)}{dz}-\Omega_{k}(0)\left(\frac{1+z}{E(z)}\right)^{2}\right)\left[1-\Omega_{k}(0)\left(\frac{1+z}{E(z)}\right)^{2}\right]^{-1},
\label{eq:omega0}
\end{equation}
being $r(z)$ the coincidence parameter which is defined as the ratio between the energy densities for dark matter and dark energy, $r = \rho_{DM}/\rho_{DE}$. The previous equation can be written in terms of the deceleration parameter if we use its standard definition, $1+q(z) = (1+z)(d\ln E(z)/dz)$, yielding
\begin{equation}
1+\frac{\omega_{DE}(z)}{1+r(z)} = \frac{2}{3}\left(1+q(z)-\Omega_{k}(0)\left(\frac{1+z}{E(z)}\right)^{2}\right)\left[1-\Omega_{k}(0)\left(\frac{1+z}{E(z)}\right)^{2}\right]^{-1},
\label{eq:omega}
\end{equation}
then, from the evaluation at present time of the previous equation, we can have an estimation for the current value of the deceleration parameter 
\begin{equation}
q_{0} = \frac{1}{2}\left(1+\frac{3 \omega_{DE,0}}{1+r_{0}}\right)\left(1-\Omega_{k}(0)\right).
\label{eq:decel}
\end{equation} 
If we consider the expression (\ref{eq:omega}) together with the values of the coincidence and curvature parameters constrained in Refs. \cite{observations, observations1} and the values obtained in \cite{decel, decel1} for the deceleration parameter\footnote{In Ref. \cite{decel} the deceleration parameter was fitted in the framework of a dark energy model. On the other hand, in Ref. \cite{decel1} some fitted values for $q$ were obtained when some parametrizations satisfying the second law of thermodynamics were considered.}, we have that $\omega_{DE,0} \in [-1.4746, -0.766296]$, which corresponds to a phantom-quintessence behavior. It is necessary to point out that if we consider a null value for the curvature parameter, $\Omega_{k}(0)$, we obtain a similar range of values for the equation of state parameter, $\omega_{DE,0}$.\\ 
If we now proceed in the opposite direction we can determine the effect of the cosmological parameters on the curvature parameter value, in Ref. \cite{des} the $\omega$-parameter was constrained to the interval $[-1.3, -0.56]$ for some dark energy models, using this interval and Eq. (\ref{eq:decel}) together with Refs. \cite{decel, decel1} for some values of the deceleration parameter, we obtain that the curvature parameter $\Omega_{k}(0)$ becomes smaller as the $\omega$-parameter decreases (phantom zone). For instance, using $\omega_{DE,0} = -1.3$ and $q_{0} = -0.57$ (see Ref. \cite{decel}), we obtain $\Omega_{k}(0) = 0.317819$, which caracterizes an open universe ($k=-1$). We can compare this result with the one obtained in Ref. \cite{omega} where $\Omega_{k}(0) \sim 0.4$ for a dynamical dark energy model with hyperbolic geometry ($k=-1$). 

\subsection{Holographic cut-off for dark energy}
\label{sec:cut}
Now we will consider the holographic principle to the dark energy problem. The physical quantities inside the universe, such as the energy density of dark energy can be described by quantities defined on the boundary of the universe. To construct $\rho_{DE}$ we will consider only the cosmological length scale, $L$, \cite{holo}. A common choice for the expression of this characteristic length is given by the Hubble scale, $L = 1/H$,
\begin{equation}
\rho_{DE} = 3c^{2}H^{2}_{0}E^{2}(z),
\label{eq:holode}
\end{equation} 
where $c$ is a positive constant in order to describe an expanding universe and it is given in the interval $0 < c^{2} < 1$. This parameter has an important role to describe the behavior of the holographic dark energy, according to its value the holographic dark energy can provide a cosmic expansion similar to the cosmological constant or the corresponding to a eternal expansion. This specific form for $\rho_{DE}$ provides an energy density similar to the dark energy present day value \cite{holo1}. Additionally, in Ref. \cite{mot1} was shown that under the election of an energy density as the one given in Eq. (\ref{eq:holode}), the second law of thermodynamics can be preserved in a (non-)flat universe. Besides, this selection for $\rho_{DE}$ had a good fit for the type Ia supernova data, as exhibited in Ref. \cite{mot2}.\\
 
Using the Eqs. (\ref{eq:fried}) and (\ref{eq:holode}) it is possible to find the energy density for dark matter,
\begin{equation}
\rho_{DM} = 3H^{2}_{0}E^{2}(z)\left[1-c^{2}-\Omega_{k}(0)\left(\frac{1+z}{E(z)}\right)^{2}\right],
\label{eq:holodm}
\end{equation}  
therefore the coincidence parameter given as the ratio $\rho_{DM}/\rho_{DE}$ can be written as
\begin{equation}
r(z) = \frac{1}{c^{2}}\left[1-c^{2}-\Omega_{k}(0)\left(\frac{1+z}{E(z)}\right)^{2}\right],
\label{eq:coincidence}
\end{equation}
and from this last expression and the values provided in Refs. \cite{observations, observations1} for the curvature and coincidence parameters, $c^{2}$ is constrained to the interval $[0.681476, 0.700786]$ up to this day. Besides, the continuity equation (\ref{eq:cont2}) becomes
\begin{equation}
(1+z)\frac{d \ln E^{2}(z)}{dz} = 3-\frac{1}{1-c^{2}}\left[\Omega_{k}(0)\left(\frac{1+z}{E(z)}\right)^{2}+\frac{Q}{3H^{3}_{0}E^{3}(z)}\right],
\label{eq:decel2}
\end{equation}
where the Eq. (\ref{eq:holodm}) was used, therefore the expression (\ref{eq:omega0}) can be rewritten as follows
\begin{eqnarray}
\frac{Q(z)}{9(1-c^{2})H^{3}_{0}E^{3}(z)} = 1-\Omega_{k}(0)\left(\frac{1+z}{E(z)}\right)^{2}\left(\frac{3-2c^{2}}{3(1-c^{2})}\right)&-&\left(1+\frac{\omega_{DE}(z)}{1+r(z)}\right)\times \nonumber \\ &\times & \left[1-\Omega_{k}(0)\left(\frac{1+z}{E(z)}\right)^{2}\right].
\label{eq:omega1}
\end{eqnarray}
At present time this equation becomes
\begin{equation}
Q_{0} = 9(1-c^{2})-3\Omega_{k}(0)(3-2c^{2})-9(1-c^{2})(1-\Omega_{k}(0))\left(1+\frac{\omega_{DE,0}}{1+r_{0}}\right).
\label{eq:Q}
\end{equation}
With the condition $Q_{0} > 0$ we have that energy flows from dark energy to dark matter sector\footnote{And vice versa for $Q < 0$.}, the coincidence problem is diminished and we are in agreement with observational data \cite{Q, Q1}. As observed, the $Q$-term written in Eq. (\ref{eq:omega1}) is constructed with the parameters of the model, i.e., we did not assume a specific parametrization for this term as is usually done. This construction may have the advantage of constraining the interaction term by using the best fit of the cosmological parameters involved, see for instance Ref. \cite{arevalo}. It is important to state that the value of $Q$ determines the rate at which the coincidence parameter decreases as the universe expands. From Eqs. (\ref{eq:fried})-(\ref{eq:cont2}) and (\ref{eq:holode}) we can write
\begin{equation}
\frac{\dot{r}}{r} = 3H(z)\omega_{DE}(z) + Q(z) \left(\frac{3H^{2}(z)-\Omega_{k}(z)}{3c^{2}H^{2}(z)\rho_{DM}(z)}\right),
\label{eq:rate}
\end{equation}
where the dot denotes a cosmic time derivative, as the curvature parameter increases or decreases the rate of change for the coincidence parameter can be altered. By considering $Q = 0$ together with $\omega_{DE} = -1$ in the above equation, we can recover the $\Lambda$-CDM model where $\dot{r} = -3Hr$. On the other hand, if we solve the condition $\dot{r} = 0$ coming from Eq. (\ref{eq:rate}) at present time we have\footnote{For simplicity in the notation for this expression we used $\Omega_{k,0}$ instead $\Omega_{k}(0)$.}
\begin{align}
& r_{0} =  \left\lbrace 3H^{2}_{0}\left[c^{2}\Omega_{k,0}+3\omega_{DE,0}(1-c^{2})(1-\Omega_{k,0})\right]  - c^{2}\left[3\rho_{DM,0}\omega_{DE,0}+\Omega_{k,0}(\Omega_{k,0}-3\omega_{DE,0}(1-\Omega_{k,0}))\right]\right. \nonumber \\
& \left. - 3\Omega_{k,0}\omega_{DE,0}(1-\Omega_{k,0})\right\rbrace \left\lbrace c^{2}\left[3\rho_{DM,0}\omega_{DE,0}-\Omega_{k,0}(3H^{2}_{0}-\Omega_{k,0})\right] \right\rbrace^{-1}, 
\end{align}
where the Eq. (\ref{eq:Q}) was used. Using the interval obtained previously for $c^{2}$ together with the values for the cosmological parameters of Refs. \cite{observations, observations1} and \cite{des} for $\omega_{DE,0}$, we can find $r_{0} \in [-373.916, 324.652]$, however, the value of the coincidence parameter to this day is highly sensitive to variations of the curvature parameter value. If we consider the value found in Ref. \cite{omega} for a dark energy model which is $\Omega_{k}(0) \sim 0.4$ ($k=-1$), we can find $r_{0} \in [-0.123364, 1.22752]$.\\ 

From the positivity condition for $Q_{0}$ and Eq. (\ref{eq:Q}) one gets
\begin{equation}
\omega_{DE,0} < (1+r_{0})\left[-\frac{c^{2}\Omega_{k}(0)}{3(1-c^{2})(1-\Omega_{k}(0))}\right],
\end{equation}
and given the previous condition for $\omega_{DE,0}$ we will have $Q_{0} > 0$ every time $\omega_{DE,0} < -0.0051394$ or $\omega_{DE,0} < 0.00567096$, where $\Omega _{k}\left( 0\right)
= 0.000_{-0.005\left(k = 1\right)}^{+0.005(k = -1)}$ in each case\footnote{If we consider the value $\Omega_{k}(0) = 0.45$ (see Ref. \cite{omega}) for an open universe, we find that for a positive $Q$-term we must have $\omega_{DE,0} < -0.836788$.} \cite{observations, observations1}. Note that when $k = 1$ the parameter $\omega_{DE,0}$ can take positive values which could lead to a decelerated expansion. From Eq. (\ref{eq:decel2}) we can write the deceleration parameter as follows:
\begin{equation}
q(z) = \frac{1}{2}\left(1-\frac{1}{1-c^{2}}\left[\Omega_{k}(0)\left(\frac{1+z}{E(z)}\right)^{2}+\frac{Q}{3H^{3}_{0}E^{3}(z)}\right]\right),
\label{eq:decelQ}
\end{equation} 
and without loss of generality, at present time
\begin{equation}
\mbox{for} \ \ Q_{0} > 0 \ \ \mbox{we \ must \ have} \ \ q_{0} < \frac{1}{2}.
\end{equation}
For a change in the direction of the energy flow, from Eq. (\ref{eq:decelQ}) and $q(z) > 1/2$ one gets
\begin{equation}
Q(z) < -3H^{3}_{0}E(z)\Omega_{k}(0)(1+z)^{2}.
\end{equation}
It is important to note that these changes in the sign of the $Q$-term interaction depend strongly on the sign of the curvature parameter. Additionally, a change in the sign of $Q$ could provide information to verify the validity of the second law of thermodynamics as well as to determine if there are phase transitions (sign changes in heat capacities) along the cosmic evolution \cite{clp, newlepe}.
 
\section{Future singularity}
\label{sec:future}
In this section we will discuss the presence of a future singularity in the model, it is important to point out that this singularity is only supported in a curved universe. In order to visualize the singularity we adopt a CPL-type parametrization for the coincidence parameter, this allows us to know the value for the redshift at which the singularity takes place. Some comments are in order. We are not dealing with a genuine Big Rip but with a Type III singularity, additionally, we can see that the cosmic evolution induced by the presence of this singularity differs from the one obtained by a cosmological constant. Using these results we show that the $Q$-term remains positive along the cosmic evolution.\\
From Eq. (\ref{eq:coincidence}) we can write the normalized Hubble parameter in terms of the coincidence parameter as follows:
\begin{equation}
E^{2}(z) = -\frac{\Omega_{k}(0)(1+z)^{2}}{c^{2}\left(r(z)-r_{c}\right)},
\label{eq:nhp}
\end{equation}
where $r_{c}$ is a constant quantity defined as $r_{c} = (1-c^{2})/c^{2}$. We can see that $E^{2}(z)$ has a singular behavior when $r(z) = r_{c}$. If we evaluate the Eq. (\ref{eq:nhp}) at present time it is possible to obtain the following expression for the constant value $r_{c}$ given in terms of some cosmological parameters
\begin{equation}
r_{c} = \frac{1-c^{2}}{c^{2}} = \frac{r_{0}+\Omega_{k}(0)}{1-\Omega_{k}(0)},
\label{eq:sing1}
\end{equation} 
where the value $c^{2} = (1-\Omega_{k}(0))/(1+r_{0})$ can be obtained from Eq. (\ref{eq:coincidence}). We will consider now a CPL-type parametrization for the coincidence parameter $r$ \cite{cpl1}
\begin{equation}
r(z) = r_{0}+\epsilon_{0}\frac{z}{1+z},
\label{eq:sing2}
\end{equation}
note that $\epsilon_{0} = r'_{0}$. We can observe that the previous parametrization becomes singular at $z=-1$, for low values of the redshift has a linear behavior and admits a bounded nature for high redshift, also owns a good sensitivity to observational data \cite{cpl2}. By equating the expressions (\ref{eq:sing1}) and (\ref{eq:sing2}) we can solve for the redshift value, $z_{s}$, at which the normalized Hubble parameter given in Eq. (\ref{eq:nhp}) becomes singular, yielding
\begin{equation}
z_{s} = -\frac{r_{0}-r_{c}}{\epsilon_{0}\left(1+(r_{0}-r_{c})/\epsilon_{0}\right)}.
\label{eq:zsing}
\end{equation}
In order to have a singular behavior in the future, we must have $-1 < z_{s} < 0$. From this condition and Eqs. (\ref{eq:sing1}) and (\ref{eq:sing2}) one gets
\begin{equation}
r(z) - r_{c} = \epsilon_{0}\left[\frac{z-z_{s}}{(1+z_{s})(1+z)} \right] \geq 0 \ \Longrightarrow \ z \geq z_{s},
\end{equation}
and from the previous equation at present time we have $(-\epsilon_{0} z_{s})/(1+z_{s}) \geq 0$, which is consistent with the condition written below Eq. (\ref{eq:zsing}). Using these results the Eq. (\ref{eq:nhp}) can be re-expressed as
\begin{equation}
E^{2}(z) = -\eta \Omega_{k}(0)\frac{(1+z)^{3}}{z-z_{s}},
\label{eq:normalized}
\end{equation}
where $\eta := (1+z_{s})/c^{2}\epsilon_{0} > 0$ since $\epsilon_{0} > 0$, as we will see later. Solving the last expression for $E^{2}(z)$, we can obtain an analytic expression for the cosmic scale factor and the redshift in terms of the cosmic time,
\begin{eqnarray}
a(t) &=& \frac{a_{0}}{1+z_{s}}\left[1-\left(\frac{3}{2}\sqrt{-\Omega_{k}(0)\eta}(1+z_{s})H_{0}(t_{s}-t) \right)^{2/3}\right], 
\label{eq:scale}
\\
1+z &=& (1+z_{s})\left[1-\left(\frac{3}{2}\sqrt{-\Omega_{k}(0)\eta}(1+z_{s})H_{0}(t_{s}-t) \right)^{2/3}\right]^{-1},
\end{eqnarray}
where we have defined
\begin{equation}
t_{s} = t_{0} + H^{-1}_{0}\left(\frac{2(-z_{s})^{3/2}}{3(1+z_{s})\sqrt{-\Omega_{k}(0)\eta}}\right),
\end{equation}
where $t_{0}$ is the initial time, therefore we have a singularity at a finite value of cosmic time in the future given by $t_{s}$. It is important to mention that $t_{s}$ is sensitive to the values of the curvature parameter; as the curvature parameter decreases, the value of $t_{s}$ increases. From Eq. (\ref{eq:scale}) we can see that as $\Omega_{k}(0) \rightarrow 0$ the scale factor tends to a constant value, i.e., a static universe, one of the main problems with this kind of universe is that is in conflict with one of the most universal laws of nature, the second law of thermodynamics \cite{perlova}, therefore under this holographic description a singular universe with a negative curvature parameter is more favored.\\ 

Note that as $t \rightarrow t_{s}$ the scale factor remains bounded $a(t\rightarrow t_{s}) \rightarrow a_{0}/(1+z_{s})$, but its first and second derivatives diverge. Additionally, since the quantity $E^{2}(z)$ can be related to the energy densities $\rho_{DE}$ and $\rho_{DM}$ through the Friedmann constraint, we have that $\rho_{DE}, \rho_{DM} \rightarrow \infty$ as $z \rightarrow z_{s}$, consequently the associated pressure to the dark energy fluid also diverges. According to this behavior we have a Type III singularity of the classification given in Refs. \cite{odintsov1, odintsov2}.
\begin{figure}[htbp!]
\centering
\includegraphics[width=7cm,height=5cm]{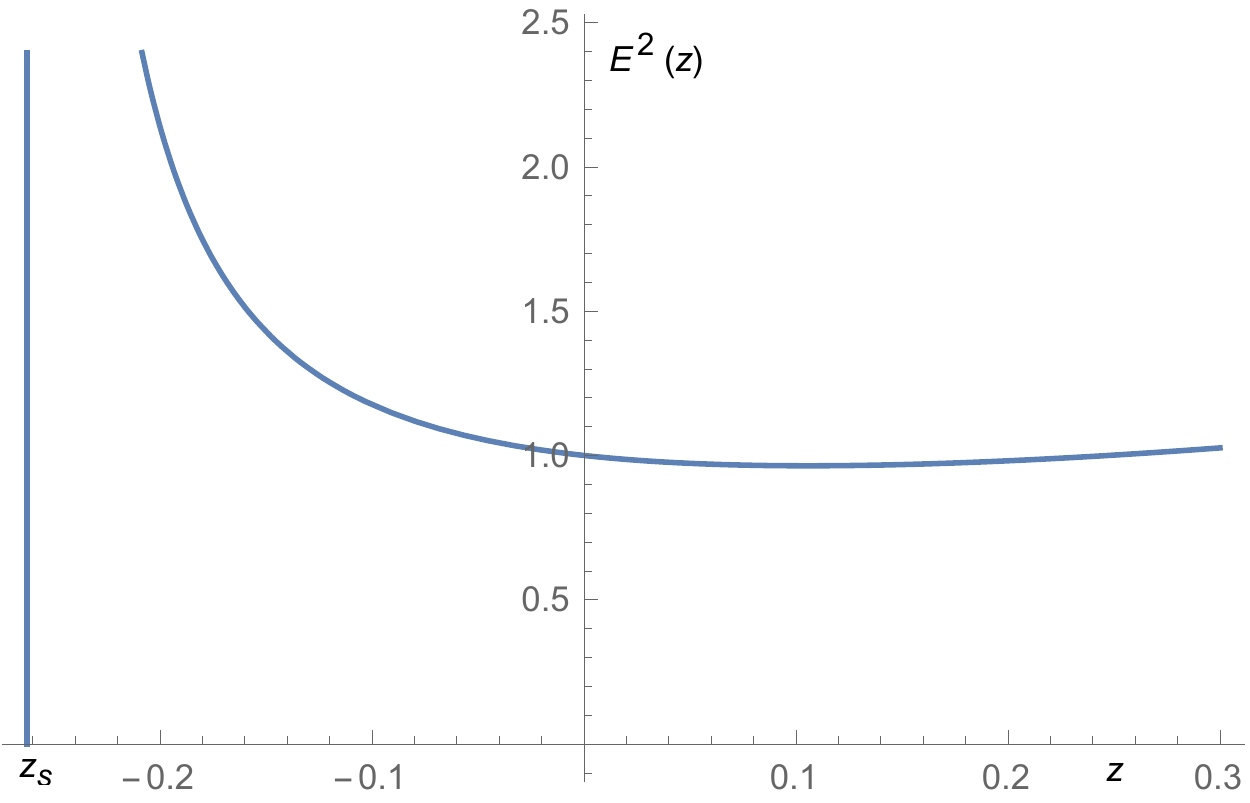}
\caption{Behavior of $E^{2}(z)$. This plot corresponds to $\Omega _{k}\left( 0\right)
= -0.005\left(k = 1\right)$.} 
\label{fig:normalized}
\end{figure}
\\

In Fig. (\ref{fig:normalized}) we can visualize the behavior of expression (\ref{eq:normalized}). Regarding this we can say that the condition $-1 < z_{s} < 0$ is not valid for any pair of values $\left\lbrace r_{0}, r_{c}\right\rbrace $. Once we consider the appropriate values for the aforementioned pair of parameters and keeping them fixed, we find that $\epsilon_{0}$ plays an important role in the manifestation of the singular behavior. As $\epsilon_{0}$ decreases, the singularity can take place close to the far future, ($z = -1$) otherwise, as $\epsilon_{0}$ increases, the singularity is closer to the present time ($z = 0$). Note that a change in the sign of the curvature parameter can cause two different types of cosmic evolution but in both cases the value $z_{s}$ is the same. In general, for other valid values of $\left\lbrace r_{0}, r_{c}\right\rbrace $ we obtain a similar behavior for $E^{2}(z)$ as observed in Fig. (\ref{fig:normalized}). As we can see, we have a crucial difference in the behavior of the normalized Hubble parameter with the one obtained in the $\Lambda$-CDM model in the far future, while in the $\Lambda$-CDM model the normalized Hubble parameter has a bounded value in our model diverges for some value of the redshift.\\

\begin{figure}[htbp!]
\centering
\includegraphics[width=7.5cm,height=6.5cm]{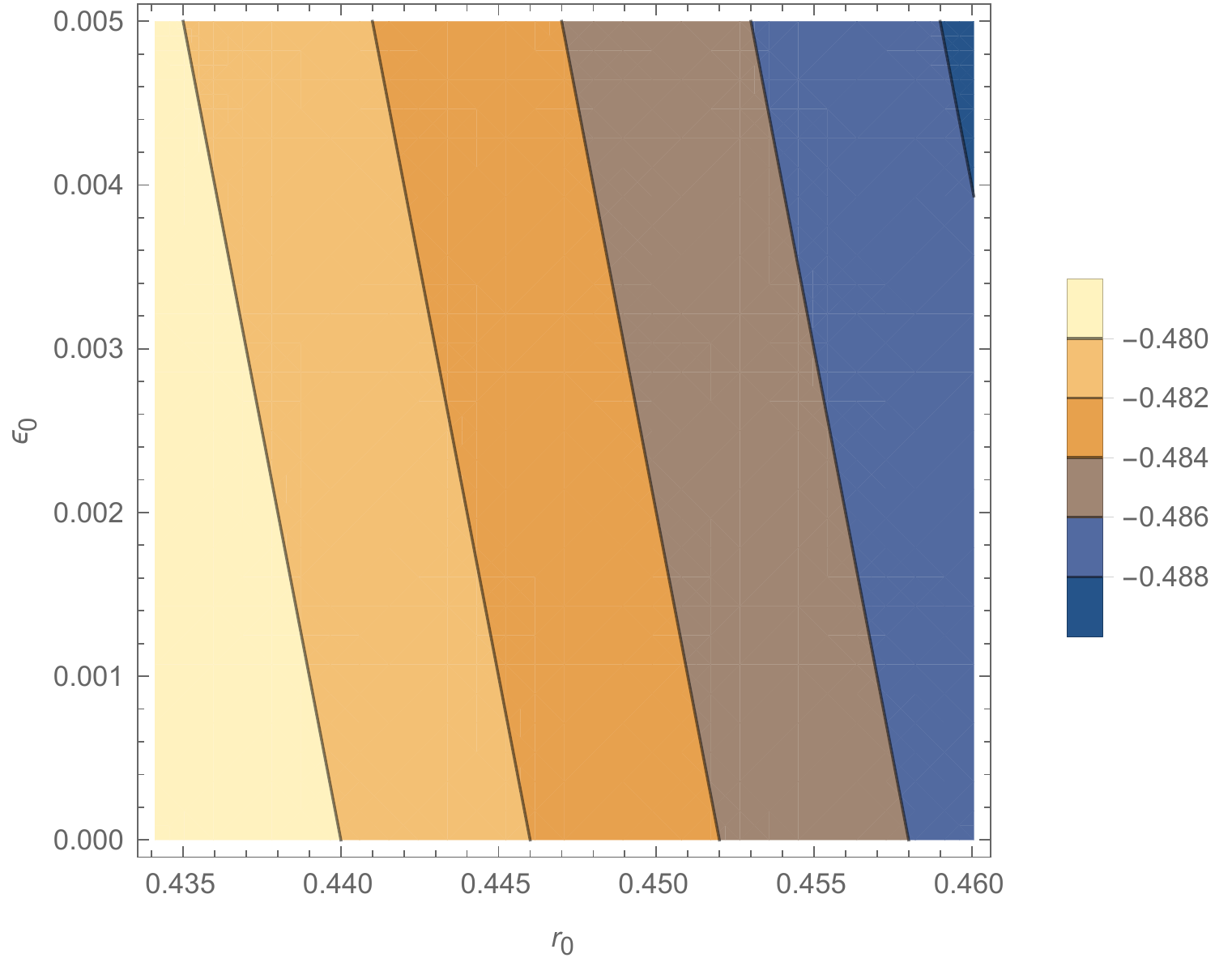}
\caption{Range of possible values for the parameter $\omega_{DE}$ in the early universe, the values for $r_{0}$ were obtained from Refs. \cite{observations, observations1}.} 
\label{fig:omegaDE}
\end{figure}

\begin{figure}[htbp!]
\centering
\includegraphics[width=7.5cm,height=6.5cm]{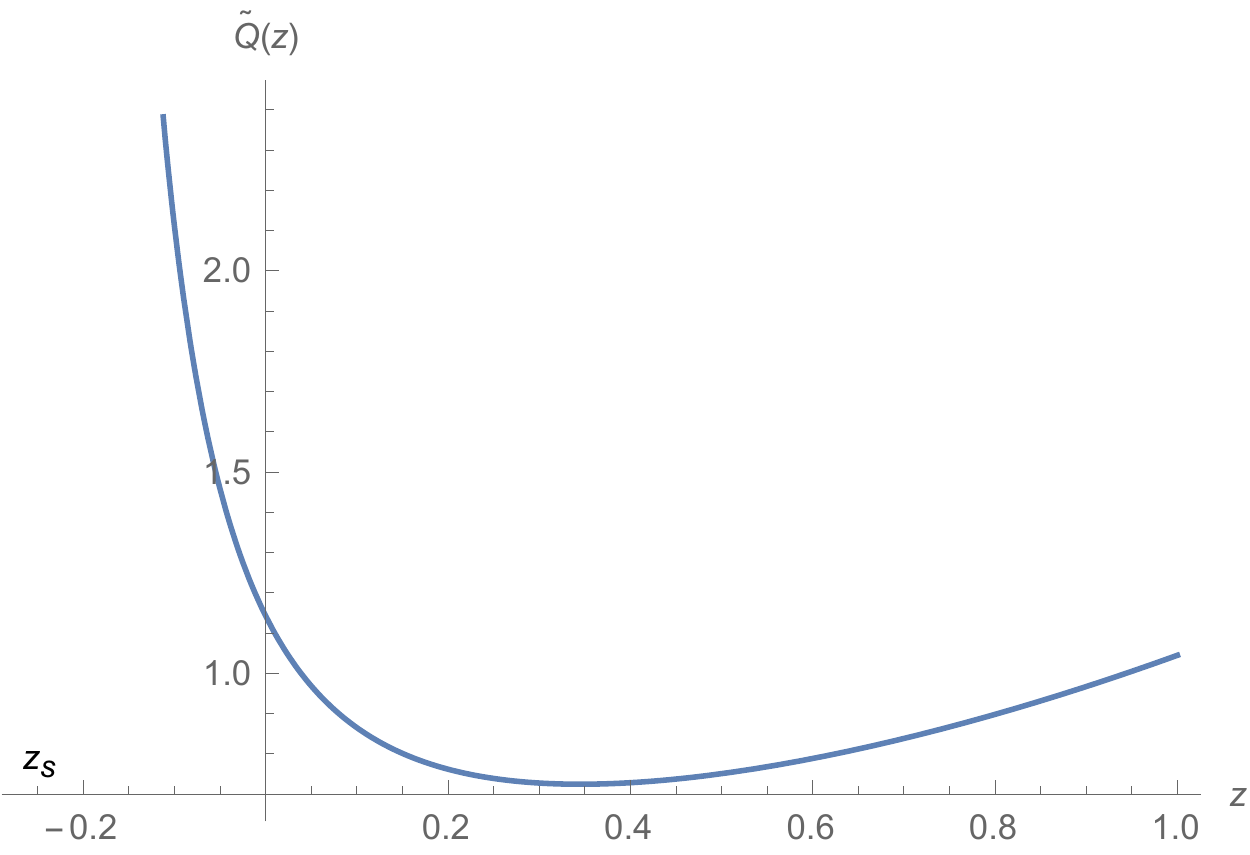}
\caption{Behavior of rescaled $Q$-term, $\bar{Q}(z) := Q(z)/3H^{3}_{0}$ with $\Omega _{k}\left( 0\right)
= -0.005\left(k = 1\right)$ and same values as in Fig. (\ref{fig:normalized}) for the pair $\left\lbrace r_{0}, r_{c}\right\rbrace$ such that the condition $-1 < z_{s} < 0$ is fulfilled.} 
\label{fig:QTerm}
\end{figure}

Moreover, if we define the function $\theta(z) := (1+z)/(z-z_{s})$ and substitute the expression (\ref{eq:normalized}) in Eq. (\ref{eq:omega0}) we get
\begin{equation}
1+\frac{\omega_{DE}(z)}{1+r(z)} = \frac{2-\eta \theta(z)(\theta(z)-3)}{3\left[1+\eta \theta(z) \right]},
\label{eq:limit}
\end{equation}
in the limit $z \rightarrow z_{s}$ we have $r(z \rightarrow z_{s}) \rightarrow r_{c}$. Therefore, from the previous equation we obtain a divergent behavior for the dark energy equation of state parameter given by $\omega_{DE}(z \rightarrow z_{s}) \rightarrow -\mbox{sgn}(1+z_{s}) \infty$, which is simply $\omega_{DE}(z \rightarrow z_{s}) \rightarrow - \infty$ since $-1 < z_{s} < 0$ for a future singularity. Otherwise, for the early universe we must consider the limit $z \rightarrow \infty$. Under this consideration the r.h.s. of Eq. (\ref{eq:limit}) tends to 2/3 and $r(z\rightarrow \infty)\rightarrow r_{0} + \epsilon_{0}$, yielding the bounded value $\omega_{DE}(z\rightarrow \infty)\rightarrow -(1+r_{0}+\epsilon_{0})/3$ (see Fig. (\ref{fig:omegaDE})). The interval given for the parameter $\epsilon_{0}$ in the plot comes from an analysis that will be detailed later. As we can observe, the parameter of equation state, $\omega_{DE}$ for the early universe takes higher values than those obtained at present time.\\
By following the same line of reasoning of the previous procedure, the expression (\ref{eq:omega1}) for the $Q$-term can be written as
\begin{equation}
\frac{Q(z)}{3H^{3}_{0}} = -\Omega_{k}(0)\sqrt{-\Omega_{k}(0)\eta \theta(z)}(1+z)^{3}\left[1+\eta \theta^{2}(z)\left(1-c^{2}\right) \right],
\label{eq:Qtheta}
\end{equation}
then, we can have an estimation of the $Q$-term for the early universe by considering the limit $Q(z \rightarrow \infty) \rightarrow \Omega_{k}(0)\left[-\mbox{sgn}\left(1+\eta \left(1-c^{2}\right)\right)\sqrt{-\mbox{sgn}\left(\Omega_{k}(0)\eta \right)}\right]\infty$. Note that the condition $Q(z \rightarrow \infty) \leq 0$ depends only on the value of the curvature parameter. However, we must consider the presence of the square root; on the other hand, near the future singularity we have $Q(z \rightarrow z_{s}) \rightarrow \Omega_{k}(0)\eta \left[(\mbox{sgn}(1+z_{s}))^{5}\mbox{sgn}(c^{2}-1)\sqrt{-\mbox{sgn}\left(\Omega_{k}(0)\eta(1+z_{s})\right)}\right]\infty$. As we will see the $Q$-term does not exhibit changes in its sign, thus the condition $Q > 0$ is kept through the cosmic evolution. This last feature is important since it can be verified by the observational data \cite{Q1, costa}. The behavior of $Q$ as a function of the redshift is depicted in Fig. (\ref{fig:QTerm}). As observed we have a monotonically increasing $Q$ function from the recent past to early times with a singular behavior at some value of the redshift in the future ($z_{s}$). If we consider other appropriate values for $\left\lbrace r_{0}, r_{c}\right\rbrace$ and $\epsilon_{0}$, we obtain similar behaviors as the one shown in Fig. (\ref{fig:QTerm}). Note that in order to have a real $Q$-function, the value of the curvature parameter plays a crucial role. Despite the observations indicate that $Q$ tends to a positive value, we must consider the possible implications that this may have at thermodynamic level for late cosmology, this is, to study the possible existence of phase transitions and the fulfillment (or not) of the second law. As shown in Ref. \cite{us}, when a future singularity is induced for a dissipative cosmology with non-linear effects, only under certain considerations the universe appears to be consistent at thermodynamical level.\\     

Additionally, from Eq. (\ref{eq:nhp}), after a straightforward calculation, we can obtain the deceleration parameter, which is given as
\begin{equation}
q(z) = -\frac{1}{2}\left(\frac{1+z}{r(z)-r_{c}}\right)r'(z),
\end{equation}
being the prime a derivative with respect to the redshift. With the use of Eqs. (\ref{eq:sing1}) and (\ref{eq:sing2}) the deceleration parameter can be rewritten as follows
\begin{equation}
q(z) = -\frac{1}{2}\left(\frac{1+z_{s}}{z-z_{s}}\right).
\label{eq:qtheta}
\end{equation}
We can observe that $q(z \rightarrow \infty) \rightarrow 0$ and $q(z \rightarrow z_{s}) \rightarrow -\infty$, i.e., the universe evolves from a non-accelerated to over-accelerated expansion. At present time we have, $\theta_{0} = -1/z_{s}$ which will be always positive, from Eqs. (\ref{eq:limit}), (\ref{eq:Qtheta}) and (\ref{eq:qtheta}), one gets
\begin{eqnarray}
\omega_{DE,0} &=& -\left(1-\frac{2-\eta \theta_{0}(\theta_{0}-3)}{3\left[1+\eta \theta_{0} \right]}\right)\left(1+r_{0}\right),\\
\frac{Q_{0}}{H^{3}_{0}} &=& -3\Omega_{k}(0)\sqrt{-\Omega_{k}(0)\eta \theta_{0}}\left[1+\eta \theta_{0}^{2}\left(1-c^{2}\right) \right],\\
q_{0} &=& -\frac{1}{2}\left(1+z_{s}\right)\theta_{0} < 0.
\end{eqnarray} 
From these expressions we can note that both conditions: positive $Q$-term at present time and cosmic evolution driven by a quintessence (phantom) fluid can be always guaranteed by fulfilling the condition $\eta \theta^{2}_{0} > -1$.\\

\begin{figure}[htbp!]
\centering
\includegraphics[width=7.8cm,height=6.8cm]{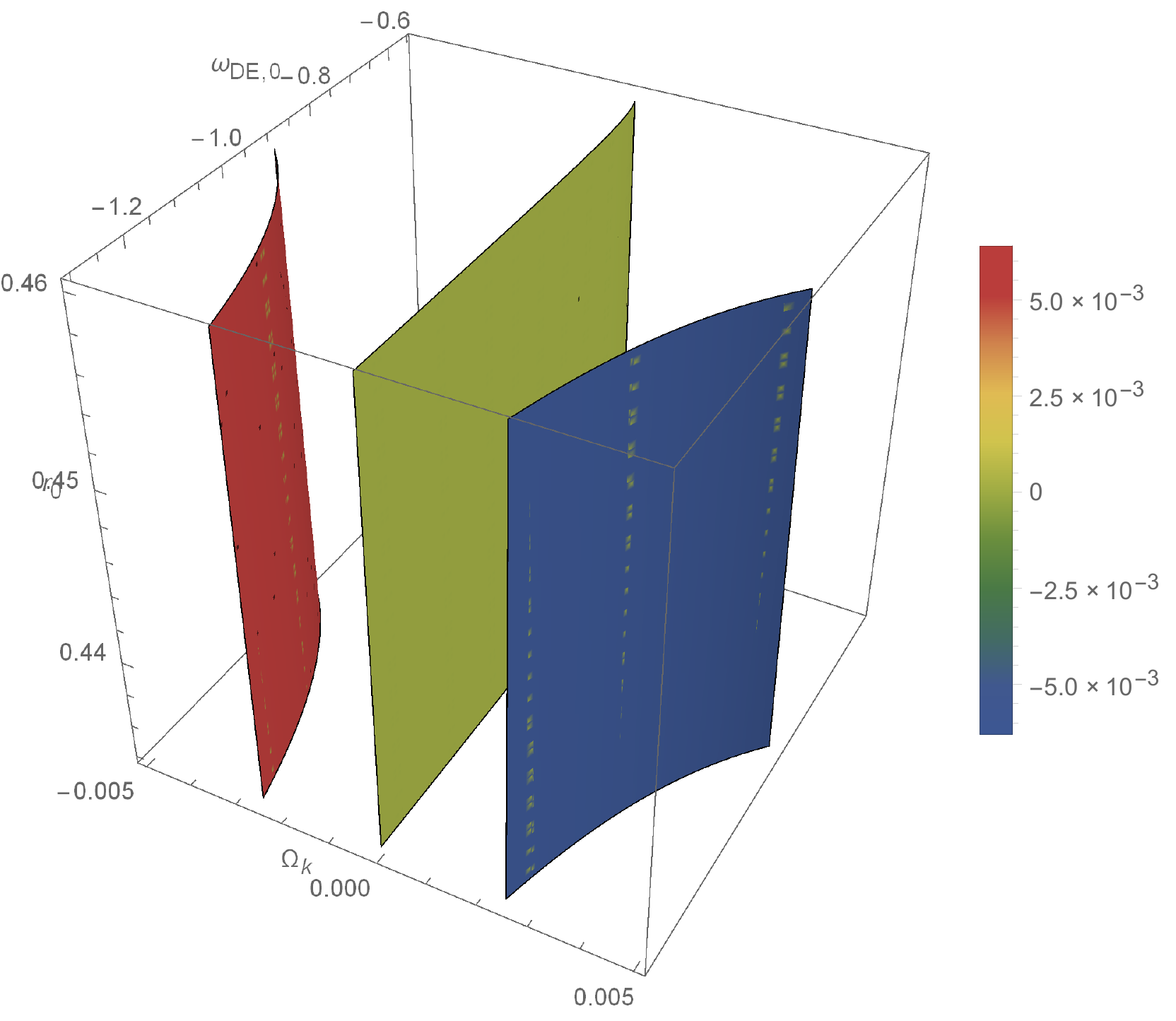}
\caption{Parameter $\epsilon_{0}$ in terms of $\omega_{DE,0}$, $r_{0}$ and $\Omega_{k}(0)$, we have considered $c^{2} = 0.681476$ which is a value within the interval determined before for $c^{2}$, for other values within this interval we obtain a similar behavior to the one shown in this plot.} 
\label{fig:eps}
\end{figure}

Finally, from the CPL parametrization given in Eq. (\ref{eq:sing2}) for the coincidence parameter and after taking its derivative with respect to the redshift, we can establish $r'_{0} := \epsilon_{0}$, therefore we have:
\begin{equation}
r'(z) = -\frac{2\Omega_{k}0}{c^{2}}\frac{(1+z)}{E^{2}(z)}\left[1-\frac{(1+z)}{E(z)}E'(z)\right],
\end{equation}
such that at present time $\epsilon_{0} = -2\Omega_{k}(0)[1-E'_{0}]/c^{2}$. Then, by using the Eqs. (\ref{eq:decel2}) and (\ref{eq:omega1}) in the previous result we can have a specific expression for $E'(z)$ at present time, i.e., 
\begin{equation}
E'_{0} = \Omega_{k}(0) + \frac{3}{2}\left(1+\frac{\omega_{DE,0}}{1+r_{0}} \right)\left(1-\Omega_{k}(0)\right).
\end{equation}
In consequence the value of $\epsilon_{0}$ can be determined as
\begin{equation}
\epsilon_{0} = -\frac{2\Omega_{k}(0)}{c^{2}}\left(1-\Omega_{k}(0)\right)\left[1 - \frac{3}{2}\left(1+\frac{\omega_{DE,0}}{1+r_{0}} \right)\right].
\end{equation} 
It is important to point out that in a similar way to the obtained throughout the analysis developed for this model, the curvature parameter plays an important role in determining an acceptable value for the parameter $\epsilon_{0}$. Note that only for closed and flat universes we can have $\epsilon_{0} > 0$, being $\Omega_{k}(0) < 0$ the most interesting case in this description. The behavior of $\epsilon_{0}$ can be seen in Fig. (\ref{fig:eps}), where we have considered \cite{observations, observations1} for the values of $r_{0}$ and the curvature parameter $\Omega_{k}(0)$ and Ref. \cite{des} for $\omega_{DE,0}$.\\

$\bullet$ {\bf Cosmological constant evolution}\\
In order to have a {\it cosmological constant} expansion we must have $q = -1$ for some value $\bar{z}$ of the redshift, then from Eq. (\ref{eq:qtheta}) we have $\bar{z} = (1+3z_{s})/2 = (1-3\left|z_{s}\right|)/2$ since $-1 < z_{s} < 0$. For instance, if we consider the value $z_{s} \sim -0.26$, which was obtained for the plots shown in the figures (\ref{fig:normalized})-(\ref{fig:statefinder2}) and (\ref{fig:QTerm}) we get $\bar{z} \sim 0.105$, therefore the cosmological constant effects were relevant in the recent past, this is consistent with what is shown by the Statefinder diagnosis in Fig. (\ref{fig:statefinder1}). This value for $z_{s}$ is not unique, we will have a different value for any appropriate pair $\left\lbrace r_{0}, r_{c}\right\rbrace$ and $\epsilon_{0}$, but, for this model everything seems to indicate that in general we will obtain always positive values for $\bar{z}$.\\ 

\section{Statefinder diagnosis}
\label{sec:state}
In this section we carry out the Statefinder diagnosis for the model once we have found the future singularity, this diagnosis reveals that the evolution for this universe is {\it far} from the behavior that it would have if the cosmological constant was responsible of its evolution.\\ 
Now, by using the expression (\ref{eq:scale}) for the cosmic scale factor, we can compute directly the following pair of parameters
\begin{equation}
r(t) = \frac{\dddot{a}}{aH^{3}}, \ \ \ \ s(t) = \frac{r-1}{3\left(q - \frac{1}{2}\right)},
\end{equation}
which is known as {\it Statefinder pair} \cite{statefinder} and $q$ is simply the deceleration parameter defined as $-\ddot{a}/aH^{2}$. The Statefinder is a geometrical tool that helps to characterize the properties of any dark energy model and {\it how far} it is from $\Lambda$-CDM by probing the expansion dynamics of the universe through the third derivative of the scale factor.\\ 
The trajectories in the $s-r$ plane for different models exhibit different behaviors. The spatially flat $\Lambda$-CDM scenario corresponds to a fixed point in the plane given by
\begin{equation}
\left\lbrace s, r\right\rbrace_{\Lambda - CDM} = \left\lbrace 0, 1\right\rbrace.
\end{equation}       
\begin{figure}[htbp!]
\centering
\includegraphics[width=7cm,height=5cm]{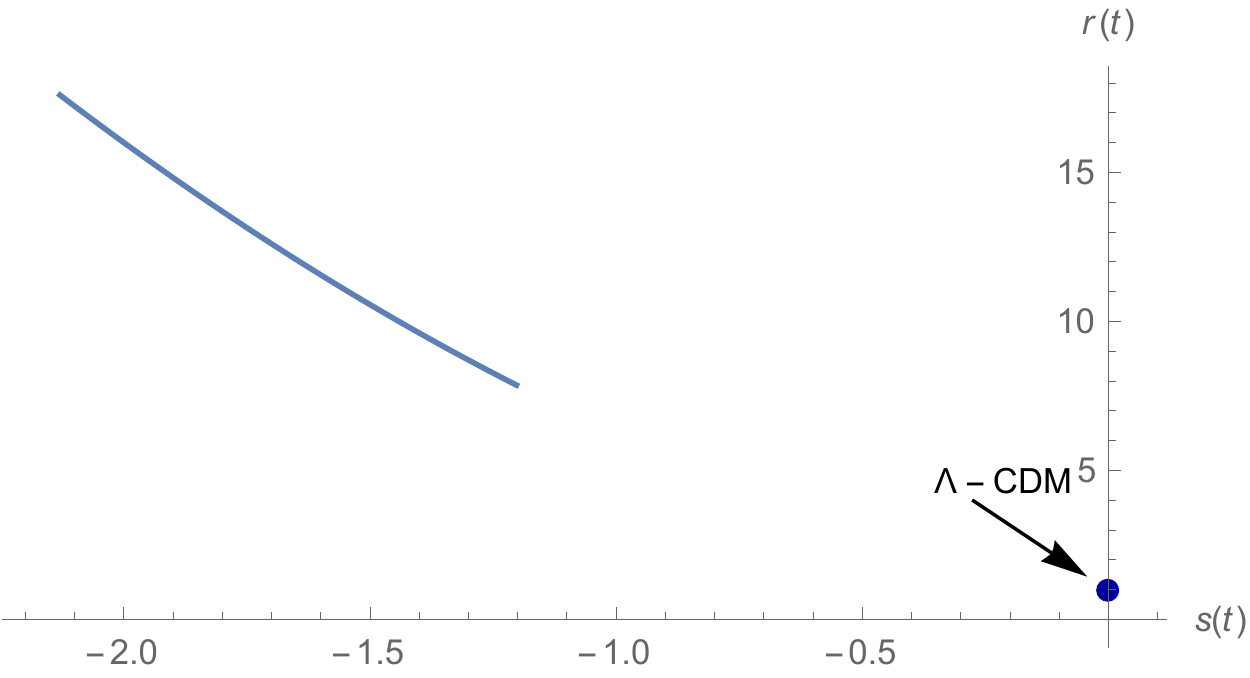}
\caption{$s(t)-r(t)$ plane for interacting dark energy model with a future singularity (phantom scenario).} 
\label{fig:statefinder1}
\end{figure}
In Fig. (\ref{fig:statefinder1}) we can visualize a trajectory in the $s-r$ plane obtained for our interacting dark energy model in the time interval $t_{0} < t < t_{s}$, for simplicity we have considered $t_{0} = 0$ together with  $\Omega _{k}\left( 0\right) = -0.005\left(k = 1\right)$, in order to have a real valued scale factor, see Eq. (\ref{eq:scale}). By keeping in mind the condition $-1 < z_{s} < 0$ for a future singularity and testing the same values for the pair $\left\lbrace r_{0}, r_{c}\right\rbrace$ as used in Fig. (\ref{fig:normalized}) (for other appropriate values of this pair we obtain similar behaviors as the one shown in the plot), we can see that due to the presence of the singular behavior, this model is {\it far} from the flat $\Lambda$-CDM model. An important feature is that the trajectory obtained in the $s-r$ plane does not contain the corresponding point to the spatially flat $\Lambda$-CDM model, this characteristic is also observed in the $q-r$ plane. From this result we can argue that under this framework the current state of the universe reveals an over-accelerated expansion which eventually will diverge. Near the singularity time $t_{s}$ the parameter $r$ explodes.\\   
\begin{figure}[htbp!]
\centering
\includegraphics[width=7.2cm,height=5cm]{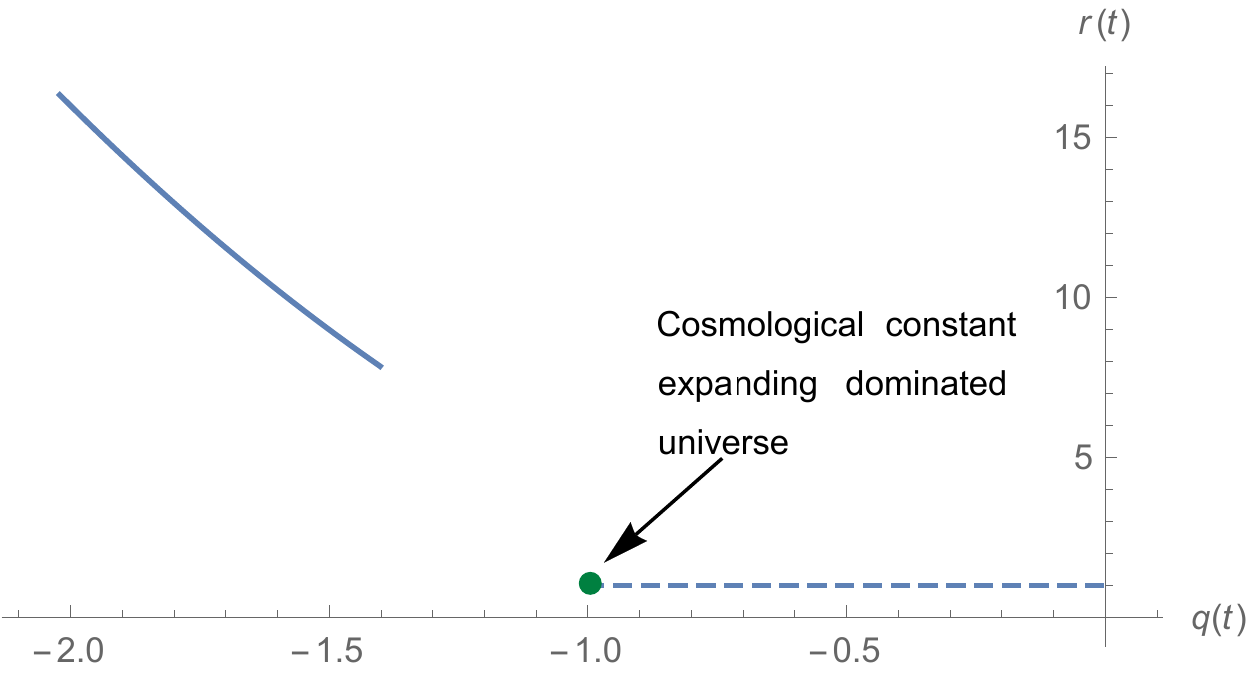}
\caption{$q(t)-r(t)$ plane.} 
\label{fig:statefinder2}
\end{figure}
On the other hand, the fact of a over-accelerated expansion can be corroborated in the $q-r$ plane (shown in Fig. (\ref{fig:statefinder2})). This model will always have negative values for the deceleration parameter (accelerated expansion). However, these values are always less than the one obtained in the $\Lambda$-CDM model ($q=-1$). In other words, the accelerated expansion in this model is driven by a phantom-fluid (over acceleration). Note that by considering a null curvature parameter in our model we obtain a static universe. Based on the above we can infer that to induce a phantom behavior in the model, we require that the value of the curvature parameter be negative ($k=1$). Despite this value for the curvature parameter characterizes a closed universe, we have that due to the presence of the future singularity the universe will not collapse as obtained in standard cosmology. 
 
\section{Final remarks}
\label{sec:final}
In this work, we present a holographic scheme for two interacting fluids in a FLRW curved spacetime, where one of these fluids represents interacting dark energy through a $Q$-term. Under this description, it was found that the curvature parameter, $\Omega_{k}(0)$ has an important role, being $\Omega_{k}(0) < 0$ the most favored. From the dynamics of the model and recent observations, can be established that the equation of state parameter for the dark energy can take values within the quintessence-phantom region at present time, i.e., the cosmic evolution has an accelerated expansion (or over-accelerated depending on the value of the $\omega$-parameter). The value for the $\omega$-parameter can change if we consider different values for the curvature parameter.\\

By considering a holographic cut-off for the dark energy density and a positive $Q$-term we can find that at present time the coincidence parameter can be constrained to a range of values which contains the one established by observations and this range can be shortened or enlarged as we vary the value of the curvature parameter. In order to maintain a positive $Q$-term, an upper bound for the $\omega$-parameter was found. However, for $\Omega_{k}(0) > 0$ the upper bound is always a positive number. This could lead to a decelerated expansion at present time. For $\Omega_{k}(0) < 0$ the upper bound is a negative number.\\

With the use of a CPL-type parametrization for the coincidence parameter, it was found that a Type III future singularity is admitted within this holographic scheme. This kind of singularity is characterized by a bounded scale factor near the singularity and divergent behavior for the densities of the fluids. In order to locate the singularity at some value of the redshift in the future ($z_{s}$), we must choose the appropriate values for the parameters involved, but such election is not unique. Therefore, the value $z_{s}$ depends on the curvature parameter. Some important quantities as the $Q$-term and the deceleration parameter were derived and both exhibit a congruent behavior near the singularity and in the early universe. As shown, the $Q$-term stays positive along the cosmic evolution, this implies that dark energy is being transformed into dark matter all the time. Some important characteristics of the model at thermodynamical level can be extracted from this interaction term, we hope to return to this point somewhere else. 
On the other hand, near the singularity the deceleration parameter approaches to $-\infty$ (over-acceleration) and in the early universe tends to a non-accelerated behavior ($q_{early} \rightarrow 0$) or $\omega_{DE,0} < \omega_{early}$. The positive constant $\epsilon_{0}$ coming from the CPL parametrization is only obtained for $\Omega_{k}(0) < 0$. With the appropriate election of the values for the parameters, we performed the Statefinder diagnosis, which showed that this singular holographic scheme is {\it far} from the $\Lambda$-CDM model, this fact can be confirmed by computing the value for the redshift from the deceleration parameter  where a cosmological constant dominated expansion holds, therefore we could establish that this phase of the universe took place in the recent past.\\

From our results we cannot perform a comparison with the flat universe since the quantities obtained are trivialized in the corresponding limit, therefore the future singularity can be obtained only in the non-flat universe within this holographic description.\\

Finally, one interesting and complementary way to strength the results showed in this work is given by testing the stability of the model but using the constructed interaction $Q$-term written in the expression (\ref{eq:omega1}), this can be performed by carrying out a perturbation analysis for this holographic model. In several works has been shown that the stability of the interacting models depends strongly on the election of the $Q$-term and the equation of state of the dark energy. In general can be found that for different choices of well known forms of the $Q$-term, the models exhibit unstable behavior under linear perturbations. We will discuss this elsewhere. 

\section*{Acknowledgments}
M.C. work has been supported by S.N.I. (CONACyT-M\'exico) and PRODEP (UV-PTC-851).

\end{document}